\documentclass[runningheads]{llncs}
\usepackage{graphicx}
\usepackage{cite}
\usepackage{url}
\usepackage{booktabs}
\usepackage{amsmath}
\usepackage{multirow}
\usepackage{tikz,xcolor,hyperref}

\definecolor{lime}{HTML}{A6CE39}
\DeclareRobustCommand{\orcidicon}{%
	\begin{tikzpicture}
	\draw[lime, fill=lime] (0,0) 
	circle [radius=0.16] 
	node[white] {{\fontfamily{qag}\selectfont \tiny ID}};
	\draw[white, fill=white] (-0.0625,0.095) 
	circle [radius=0.007];
	\end{tikzpicture}
	\hspace{-2mm}
}

\foreach \x in {A, ..., Z}{%
	\expandafter\xdef\csname orcid\x\endcsname{\noexpand\href{https://orcid.org/\csname orcidauthor\x\endcsname}{\noexpand\orcidicon}}
}

\begin{document}
%

\title{Fully automated deep learning based segmentation of normal, infarcted and edema regions from multiple cardiac MRI sequences}

\titlerunning{Deep CNN based automated CMR segmentation}
\author{Xiaoran Zhang\inst{1}\orcidA{} \and Michelle Noga\inst{2,3} \and Kumaradevan Punithakumar\inst{2,3}\orcidB{}} 

%
\authorrunning{Zhang et al.}
%
\institute{Dept. of Electrical and Computer Engineering, UCLA, Los Angeles, USA \and
Dept. of Radiology \& Diagnostic Imaging, University of Alberta, Canada \and
Servier Virtual Cardiac Centre, Mazankowski Alberta Heart Institute, Canada\\
\email{punithak@ualberta.ca}
}

\maketitle 

\begin{abstract}
Myocardial characterization is essential for patients with myocardial infarction and other myocardial diseases, and the assessment is often performed using cardiac magnetic resonance (CMR) sequences. In this study, we propose a fully automated approach using deep convolutional neural networks (CNN) for cardiac pathology segmentation, including left ventricular (LV) blood pool, right ventricular blood pool, LV normal myocardium, LV myocardial edema (ME) and LV myocardial scars (MS). The input to the network consists of three CMR sequences, namely, late gadolinium enhancement (LGE), T2 and balanced steady state free precession (bSSFP). The proposed approach utilized the data provided by the MyoPS challenge hosted by MICCAI 2020 in conjunction with STACOM. The training set for the CNN model consists of images acquired from 25 cases, and the gold standard labels are provided by trained raters and validated by radiologists. The proposed approach introduces a data augmentation module, linear encoder and decoder module and a network module to increase the number of training samples and improve the prediction accuracy for LV ME and MS. The proposed approach is evaluated by the challenge organizers with a test set including 20 cases and achieves a mean dice score of $46.8\%$ for LV MS and $55.7\%$ for LV ME+MS.

\end{abstract}

\section{Introduction}
The imaging-based assessment of the heart using modalities such as magnetic resonance imaging (MRI) plays a central role in the diagnosis of cardiac disease, a leading cause of death worldwide. Late gadolinium-enhanced (LGE) imaging is one of the commonly used cardiac magnetic resonance (CMR) sequences to diagnose myocardial infarction \cite{arai_magnetic_2011}, a common cardiac disease that may lead to heart failure. Acute injury or inflammation related to other conditions can be detected using T2-weighted CMR. However, detecting ventricular boundaries using the LGE or T2-weighted images is challenging, while this function can more easily be performed using a balanced steady state free precession (bSSFP) sequences. Often many cardiac patients are scanned using multiple CMR sequences, and utilizing the combination of these sequences will allow for obtaining robust and accurate diagnostic information \cite{zhuang2018multivariate}. 

The target of this study is to combine the multi-sequence CMR data to produce an accurate segmentation of cardiac regions including left ventricular (LV) blood pool (BP), right ventricular BP, LV normal myocardium (NM), LV myocardial edema (ME) and LV myocardial scars (MS) and specifically focus on classifying myocardial pathology. Generally, the myocardium region could be divided into normal, infarcted and edematous regions. Generating accurate contour for these regions is arduous, time-consuming and thus automating the segmentation process is of great interest \cite{ukwatta_myocardial_2016}. Zabihollahy \emph{et al.} \cite{zabihollahy2019convolutional} proposed a semiautomatic tool for LV scar segmentation using CNNs. Li \emph{et al.} \cite{li2020atrial} proposed a fully automatic tool for left atrial scar segmentation.

In this challenge, there are mainly two difficulties to produce an accurate prediction of the LV ME and MS. The first difficulty is the limited amount of training data which only consists of 25 cases. The second is the small size of the LV ME and MS regions with high intra and inter-subject variations. The inter-observer variation of manual scar segmentation is reported with a Dice score of $0.5243\pm0.1578$ \cite{zhuang2018multivariate}. 

In this study, we propose a fully automated approach by utilizing deep convolutional neural networks to delineate the LV BP, RV BP, LV NM, LV ME and LV MS regions from multi-sequence CMR data including bSSFP, LGE and T2. Our main contributions are the following: 1) we introduce a data augmentation module and increase the training size by 40 times using random warping and rotation; 2) we introduce a linear encoder and decoder to improve the network's training performance and utilize three different architectures including a shallow version of the standard U-net \cite{ronneberger2015u}, Mask-RCNN \cite{he2017mask} and U-net++\cite{zhou2018unet++,zhou2019unetplusplus} for the LV ME and LV MS block and average the predictions of the three networks followed by a binary activation with threshold 0.5. Our method is evaluated by the challenge organizers on a test set consisting of 20 cases, which contain images acquired from scanners that are not included in the training set.

\section{Methodology}
\begin{figure}[htbp]
    \centering
    \includegraphics[scale=0.435]{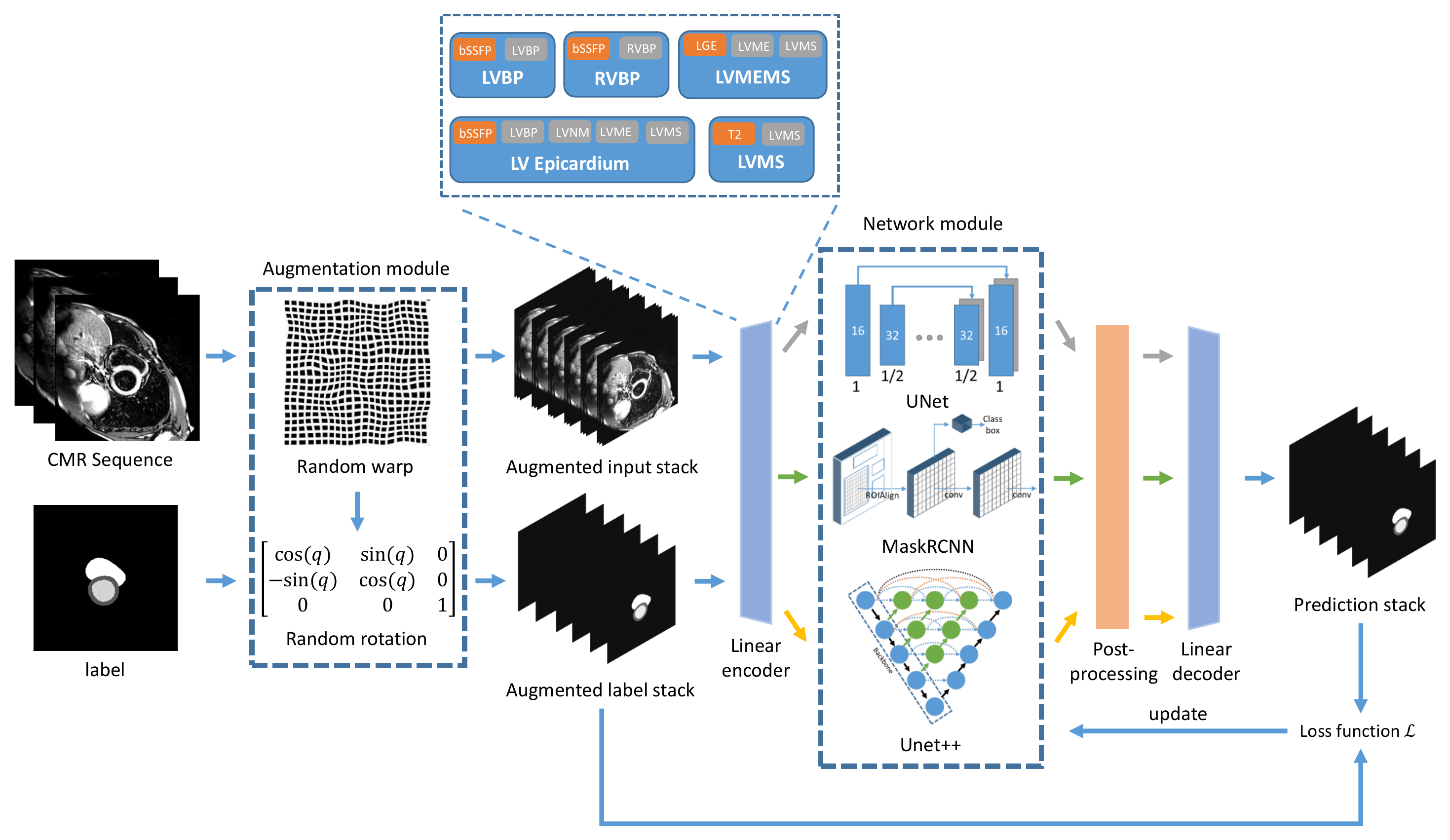}
    \caption{Overall architecture of the proposed method in the training stage}
    \label{fig:model_arch}
\end{figure}

We introduce the pipeline shown in Fig.~\ref{fig:model_arch} for the LV, ME and MS segmentation. The proposed method is fully automatic and utilized no additional data other than the training set provided by the challenge organizers. The details of each module are introduced in the following sections.

\subsection{Augmentation module}
We first extract the input data in a slice-by-slice manner and perform center cropping to obtain images of size $256\times 256$. To improve the number of samples, we perform two data augmentation schemes including random warping and random rotation. The random warping is performed by firstly generating a $8\times8\times2$ uniformly distributed random matrix, where the last dimension indicates 2D space, with each entry in range $[-5,5]$. We then resize the non-rigid warping matrix to the image size with dimension $256\times256\times2$ and apply the warping map using bilinear interpolation. The extracted input CMR slices and the labels are warped 20 times. After augmenting the data using random warping, we then utilize random rotation in $90^\circ$, $180^\circ$ and $270^\circ$ with equal probability. The training set is then augmented with one time more data with random choice among the three options. 

\subsection{Preprocessing}
All training and validation images are normalized using $5^{\textrm{th}}$ and $95^{\textrm{th}}$ percentile values, $I_{05}$ and $I_{95}$, of the intensity distribution of the 2D data to obtain relatively uniform training sets. The normalized intensity value, $I_n$, is computed using
\begin{math}
    I_n = \displaystyle\frac{I-I_{05}}{I_{95}-I_{05}}
\end{math}
where $I$ denotes the original pixel intensity.

\subsection{Linear encoder}
We introduce a linear encoder and a corresponding decoder for the augmented input stack after preprocessing. Inspired by the clinical observation in \cite{zhuang2018multivariate}, we encode the augmented input and label stacks and produce five input blocks as shown in Fig.~\ref{fig:model_arch} instead of blindly concatenating the CMR sequences, where each block represents the data used to train a target class. The five input blocks are \textbf{LVBP block}, which uses bSSFP as the input image and LV BP as the target; \textbf{RVBP block}, which uses bSSFP as the input and RV BP as the target; \textbf{LV Epicardium block}, which uses bSSFP as the input and the linear combination of LV BP, LV NM, LV ME and LV MS as the target; \textbf{LVMEMS block}, which uses LGE as the input and the combination of LV ME and LV MS as the target; and the \textbf{LVMS block}, which uses the T2 as the input and LV MS as the target. In the testing mode, the linear encoder will only perform on the input stack. The network module will infer on the encoded input and the decoder will extract the predictions after post-processing.

\subsection{Network module}
In order to improve the performance for the edema and scar prediction, we utilize three different architectures with different input blocks for each model. The results are averaged from the three networks for LV ME+MS and LV MS and followed by a binary activation with threshold 0.5. The details for each network are shown in the following. 

\subsubsection{U-net}
\begin{figure}[htbp]
    \centering
    \includegraphics[scale=0.35]{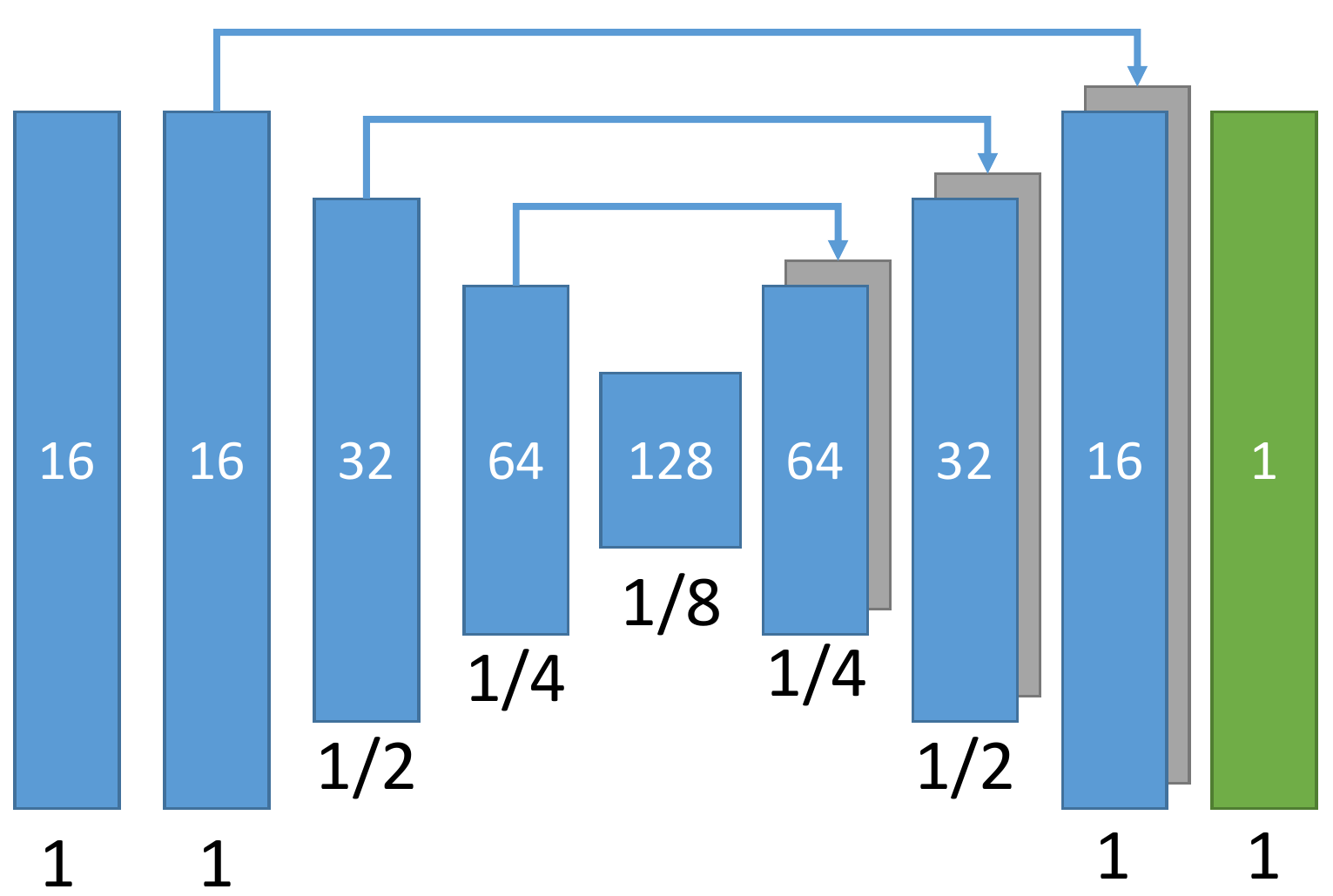}
    \caption{Architecture of the U-net model. The blue block indicates the $3\times3$ convolution layer and the number indicates channel size. The blue arrow indicates the skip connection. The green block indicates the $1\times1$ convolution layer with sigmoid activation to produce the prediction masks.}
    \label{fig:UNet}
\end{figure}

The U-net module utilizes a shallow version shown in Fig.~\ref{fig:UNet} of the standard U-net \cite{ronneberger2015u} to prevent overfitting. The U-net is trained on all the five input blocks produced by the linear encoder. The loss function of the U-net is selected as the negative of dice coefficient with Adam optimizer (learning rate = $1e-5$) and batch size = $8$. 

\subsubsection{Mask-RCNN:}
The Mask-RCNN module \cite{he2017mask} utilizes ResNet50 \cite{he2016deep} as the backbone for the segmentation task and is implemented using the Matterport library \cite{matterportmaskrcnn2017}. The Mask-RCNN is trained on the LVMEMS and LVMS using Adam optimizer (learning rate = $0.001$) and batch size = $2$.

\subsubsection{U-net++:} The U-net++ module \cite{zhou2018unet++,zhou2019unetplusplus} utilizes VGG16 \cite{simonyan2014very} as the backbone. The model is trained on the LVMEMS and LVMS using the negative of dice coefficient as the loss function with Adam optimizer (learning rate = $1\times 10^{-5}$) and batch size = $8$. 

\subsection{Post-processing}
We applied post-processing to retain only the largest connected component for the predictions of LV BP and LV Epicardium by U-net. The operation is performed in 2D space with a slice-by-slice manner. In addition, we applied an operation to remove holes that appear inside the foreground masks. As shown in Fig.~\ref{fig:model_arch}, the post-processing is performed on the encoded predictions before the linear decoder.

\subsection{Linear decoder}
The corresponding decoder performs the linear subtraction on the predicted masks of LV Epicardium and LVMEMS and is followed by a binary activation for all predicted masks in five target classes with threshold 0.5. The decoder also includes a binary constraint for the LV ME and MS predictions by calculating the myocardium mask using the predicted LV epicardium and LV BP and performing a pixelwise multiplication
\begin{equation}
   P_i=\begin{cases}
         \sigma(\tilde{P}_0) & i=0 \\
         \sigma(\tilde{P}_1) & i=1 \\
         \sigma(\tilde{P}_2 - \tilde{P}_0 - \tilde{P}_3) & i=2 \\
         \sigma(\tilde{P}_2-\tilde{P}_0)\odot\sigma(\tilde{P}_3-\tilde{P}_4) & i=3 \\
         \sigma(\tilde{P}_2-\tilde{P}_0)\odot\sigma(\tilde{P}_4) & i=4
         
    \end{cases}
\end{equation}
where $i=0,1,2,...,4$ denotes the index for LVBP block, RVBP block, LV Epicardium block, LVMEMS block, LVMS block respectively. $P_i$ denotes the final prediction mask and $\tilde{P}_i$ denotes the raw prediction after post-processing for block $i$. $\sigma(\cdot)$ denotes the binary activation function with threshold 0.5. The notation $\odot$ denotes the pixelwise multiplication.

\section{Experiments}
\subsection{Clinical data}
The training set consists of 25 cases of multi-sequence CMR and each refers to a patient with three sequence CMR including bSSFP, LGE and T2. The training data is processed using the MvMM method \cite{zhuang2016multivariate,zhuang2018multivariate}. The training set labels include LV BP (labelled 500), RV BP (600), LV NM (200), LV ME (1220) and LV MS (2221). The manual segmentation is performed by trained examiners and corrected by experienced radiologists. The test set consists of another 20 cases of multi-squence CMR and the ground truth is not provided.

\subsection{Implementation details}
The networks are implemented using Python programming language with Keras and Tensorflow. All networks are trained with 500 epochs on NVIDIA Tesla--P100 graphical processing units with 12 GB memory. The trained neural network model with the highest validation accuracy is saved to the disk. The validation split is 0.8 for all networks with 3264 images for training and 816 images for validation after the data augmentation module. The original extracted 2D slices from the training data provided by challenge organizers contain 102 images.

\subsection{Evaluation metrics}
\subsubsection{Dice coefficient}
DC measures the overlap between two delineated regions \cite{sorensen1948method}:
\begin{equation}
DC=\frac{2|A\bigcap M|}{|A|+|M|}
\end{equation}
where set $A$ as the automatic prediction region and set $M$ as manual segmentation ground truth.

\subsubsection{Jaccard index}
Jaccard index measures the similarity and diversity between two delineated regions \cite{jaccard1912distribution}:
\begin{equation}
J=\frac{|A\bigcap M|}{|A|+|M|-|A\bigcap M|}.
\end{equation}

\section{Results}
The proposed method is evaluated over images acquired from a total of 20 cases including CMR sequences consists of bSSFP, LGE and T2. The evaluation of the proposed method on test sets are performed by the challenge organizers with dice score on LV ME+MS and LV MS. The ground truth of the test sets are not shared with the participants. 

\subsection{Quantitative assessment}


\begin{table}[htbp]
	\centering
	\setlength{\tabcolsep}{10pt}
	\caption{Overall performance of UNet, proposed method without the linear encoder and decoder, and the proposed method evaluated over CMR test datasets acquired from 20 cases on LV ME+MS and LV MS.}	
	\begin{tabular}{lcccc}
		\toprule
		\multirow{2}{*}{Methods} & \multicolumn{2}{c}{Dice metric (\%)} & \multicolumn{2}{c}{Jaccard index(\%)} \\
		                         &  MS                  & ME+MS                 & MS                & ME+MS                         \\
		\midrule
		UNet         & $36.2\pm23.2$          & $43.2\pm16.0$           & $24.5\pm18.1$       & $28.8\pm13.1$\\
		Proposed$^\dagger$      & $38.5\pm24.3$         & $54.2\pm17.1$            & $26.5\pm18.9$      & $38.9\pm15.0$\\
		Proposed                & $\mathbf{46.8\pm 26.8}$        & $\mathbf{55.7\pm 18.3}$        & $\mathbf{34.2\pm 22.2}$   & $\mathbf{40.5\pm 16.3}$\\
		\bottomrule
		\multicolumn{5}{l}{$^\dagger$ indicates without the linear encoder and decoder module.}
	\end{tabular}
	\label{tbl:overall}
\end{table}
The agreement between the segmentation of the proposed approach with the manual ground truth is quantitatively evaluated using the dice metric and Jaccard index. 
To illustrate the effectiveness of the network module and the linear encoder and decoder, we report the performance of the UNet, proposed method without the linear encoder and decoder, and the proposed method in Table.~\ref{tbl:overall}. The best result for the test set achieves a mean dice score of $46.8\%$ for LV MS and $55.7\%$ for LV ME+MS. Our proposed network module improves the overall performance of MS and ME+MS by comparing our proposed method without the linear encoder and decoder with the UNet. Our proposed linear encoder and decoder module further improves the performance especially in the MS segmentation. Fig.~\ref{fig:box} shows the performance of the three methods using box plots.
\begin{figure}[htbp]
    \centering
    \setlength{\tabcolsep}{2pt}
    {\scriptsize
    \begin{tabular}{cc}
     \includegraphics[scale=0.32]{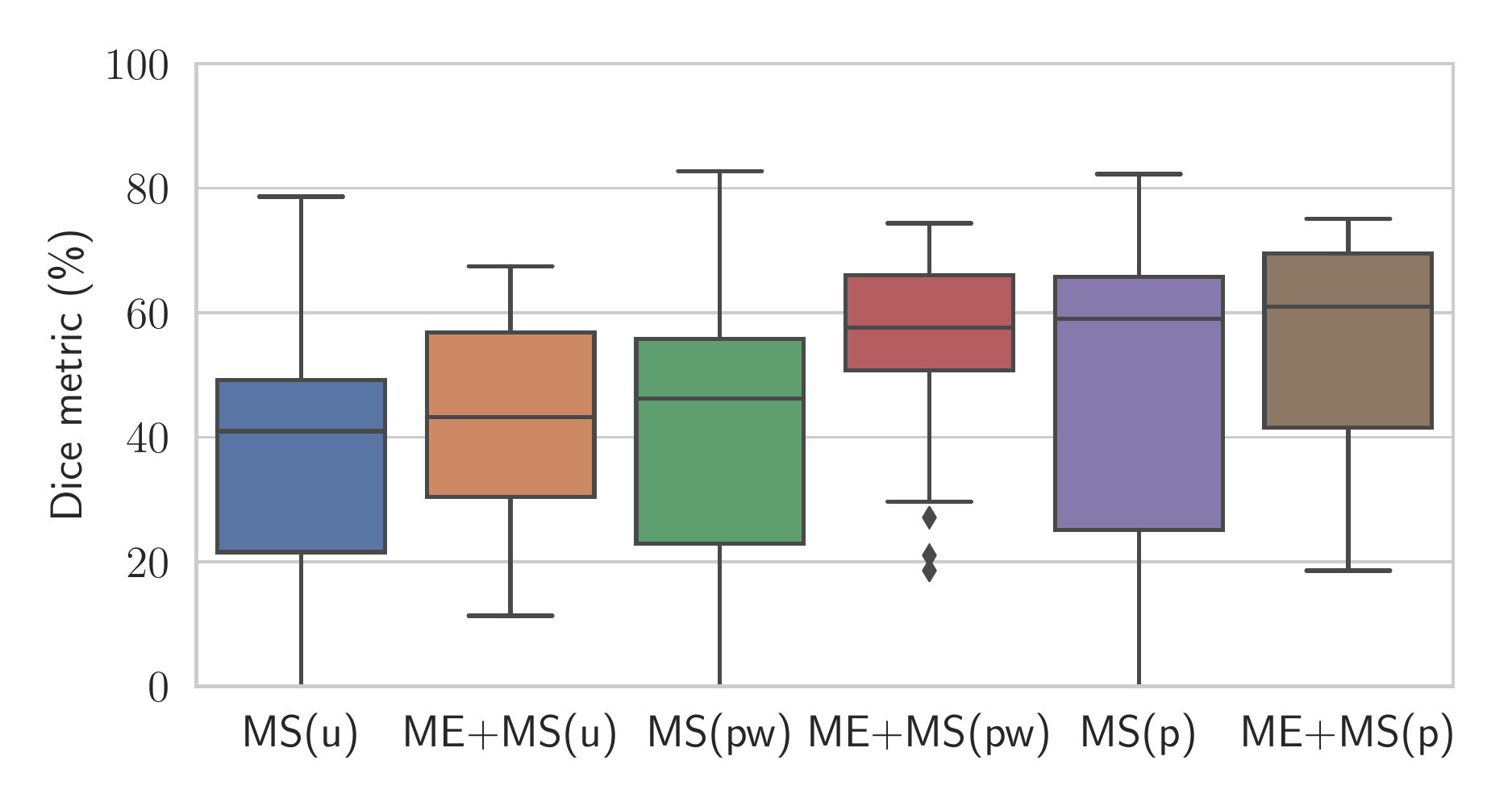}    &
     \includegraphics[scale=0.32]{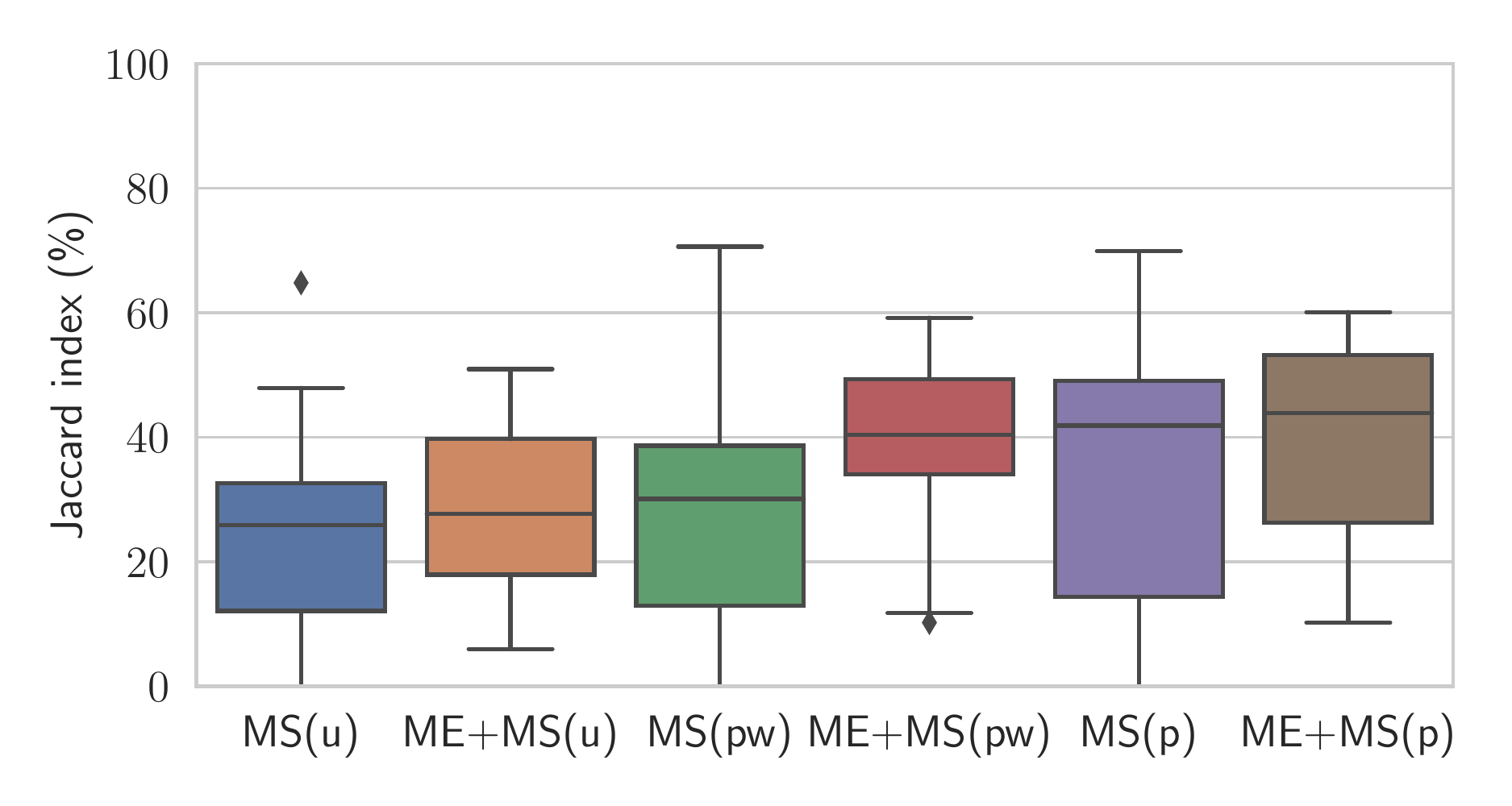} 
    \end{tabular}}
    \caption{The boxplots showing the performance of the proposed approach over test sets acquired from 20 cases. The evaluations were performed using Dice metric and Jaccard index. In the figure, $u$ indicates the results by UNet; $pw$ indicates the proposed method without linear encoder and decoder; $p$ indicates the proposed method.}
    \label{fig:box}
\end{figure}

\subsection{Visual assessment}
\begin{figure}[h!]
    \centering
    \setlength{\tabcolsep}{1pt}
    {\scriptsize
    \begin{tabular}{ccccc}
         \includegraphics[width=0.195\columnwidth, trim=1.8cm 1.8cm 1.8cm 2.5cm, clip]{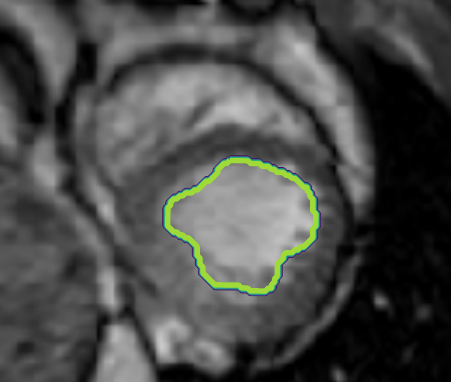} &
         \includegraphics[width=0.195\columnwidth, trim=1.8cm 1.8cm 1.8cm 2.5cm, clip]{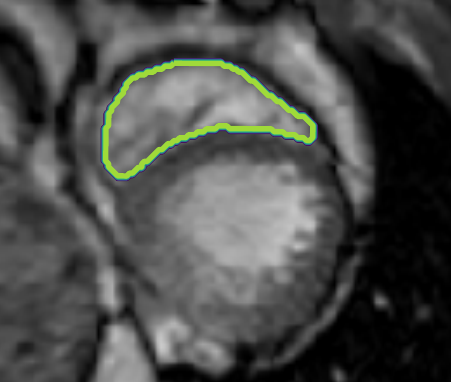} &
         \includegraphics[width=0.195\columnwidth, trim=1.8cm 1.8cm 1.8cm 2.5cm, clip]{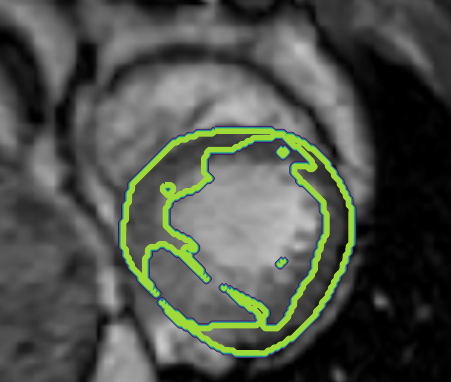} &
         \includegraphics[width=0.195\columnwidth, trim=1.8cm 1.8cm 1.8cm 2.5cm, clip]{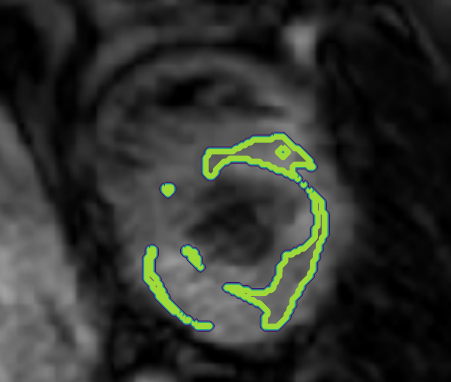} &
         \includegraphics[width=0.195\columnwidth, trim=1.8cm 1.8cm 1.8cm 2.5cm, clip]{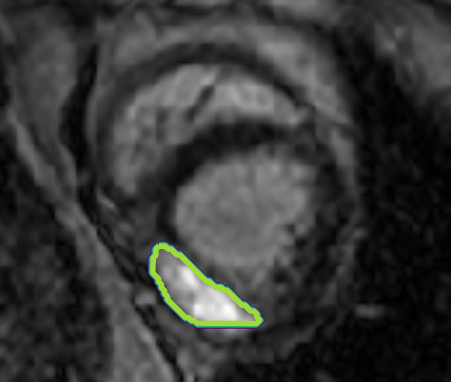} \\
         \includegraphics[width=0.195\columnwidth, trim=1.5cm 1.3cm 1.5cm 1.8cm, clip]{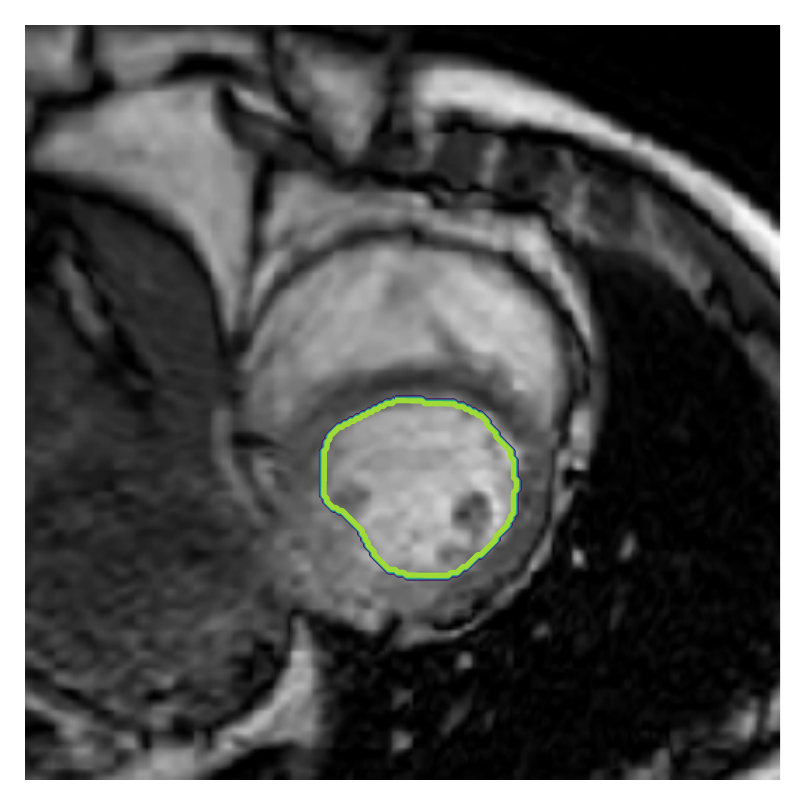} &
         \includegraphics[width=0.195\columnwidth, trim=1.5cm 1.3cm 1.5cm 1.8cm, clip]{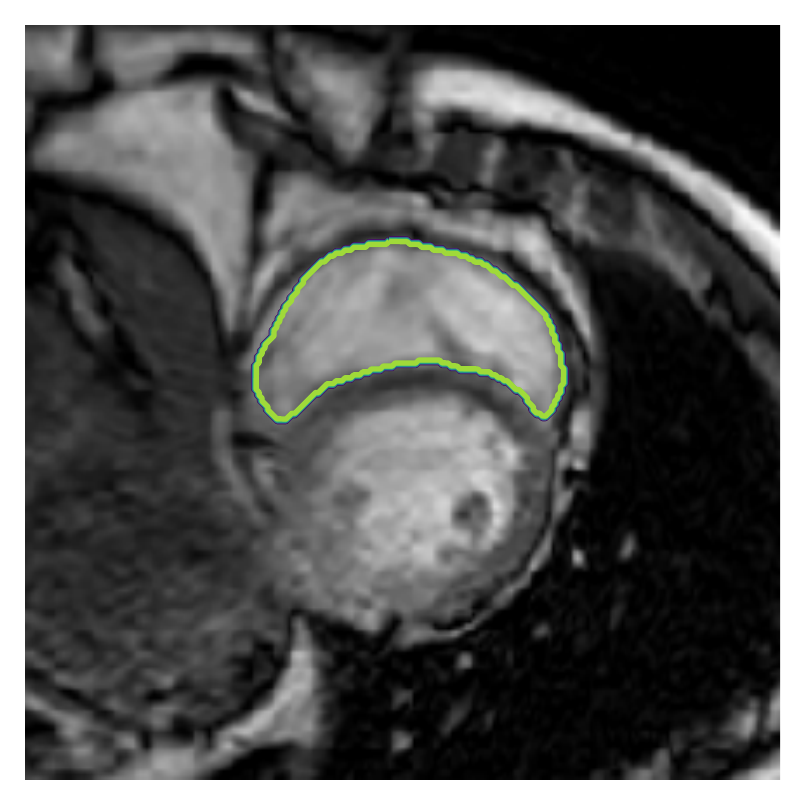} &
         \includegraphics[width=0.195\columnwidth, trim=1.5cm 1.3cm 1.5cm 1.8cm, clip]{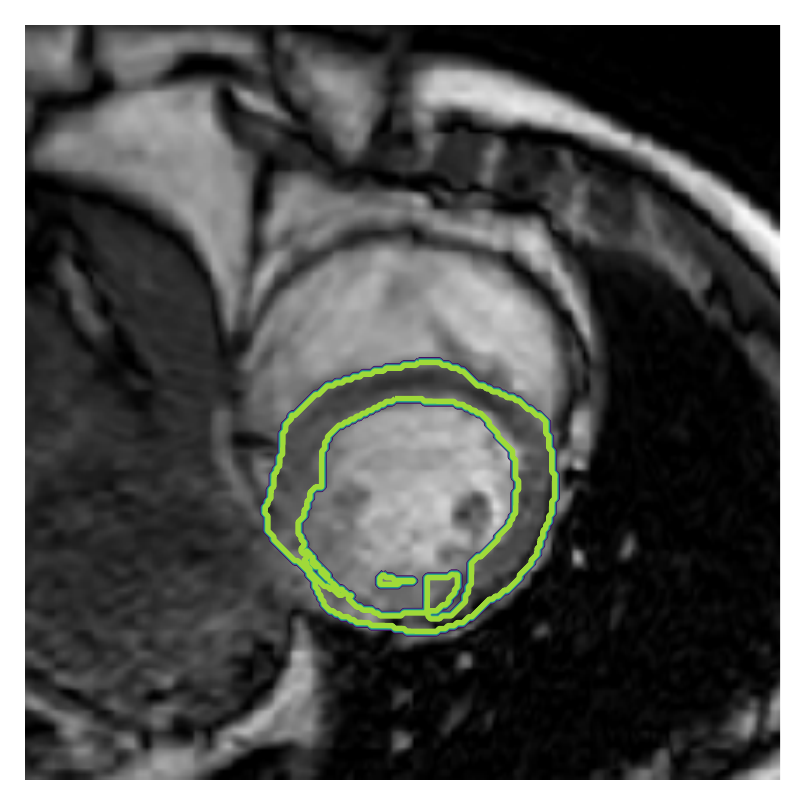} &
         \includegraphics[width=0.195\columnwidth, trim=1.5cm 1.3cm 1.5cm 1.8cm, clip]{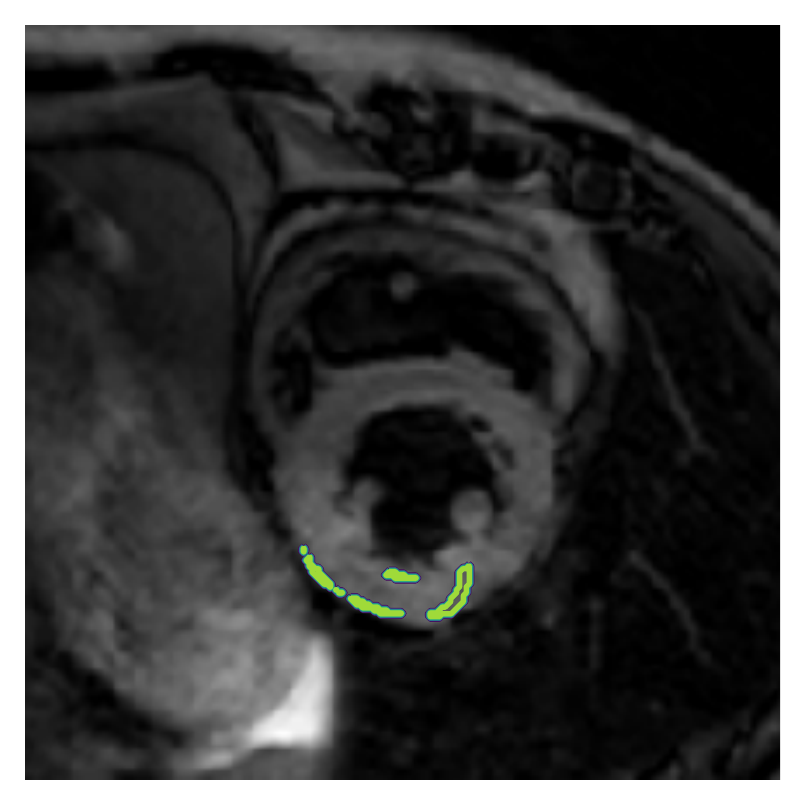} &
         \includegraphics[width=0.195\columnwidth, trim=1.5cm 1.3cm 1.5cm 1.8cm, clip]{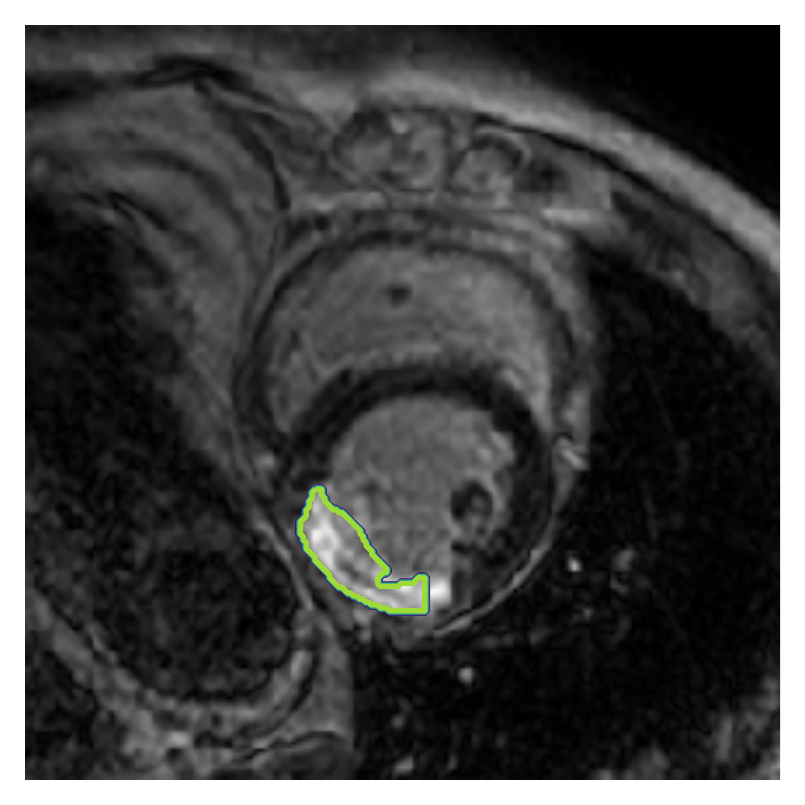} \\
         \includegraphics[width=0.195\columnwidth, trim=1.7cm 1.3cm 1.3cm 1.7cm, clip]{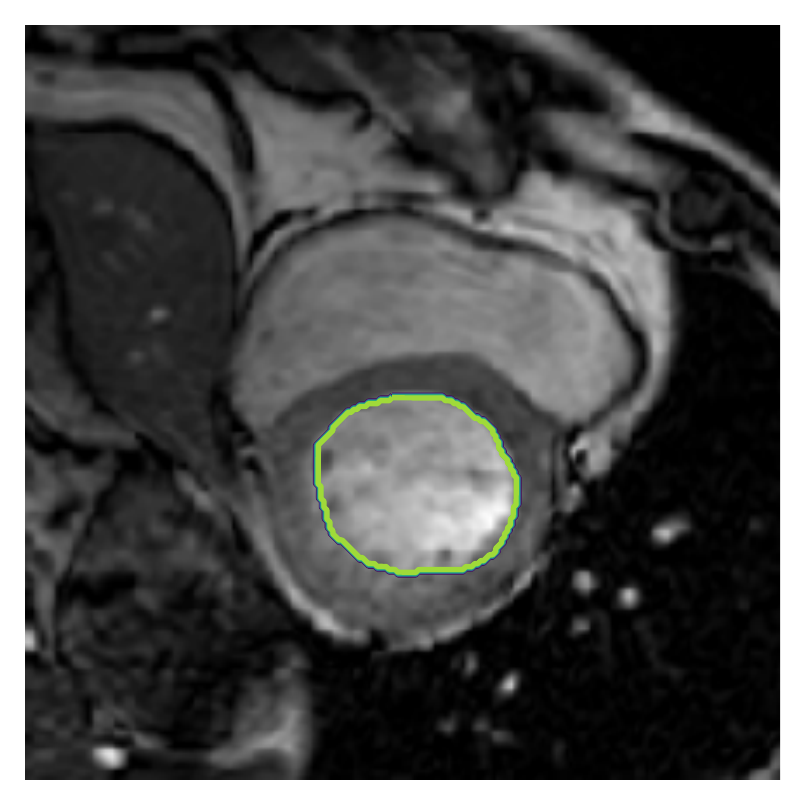} &
         \includegraphics[width=0.195\columnwidth, trim=1.7cm 1.3cm 1.3cm 1.7cm, clip]{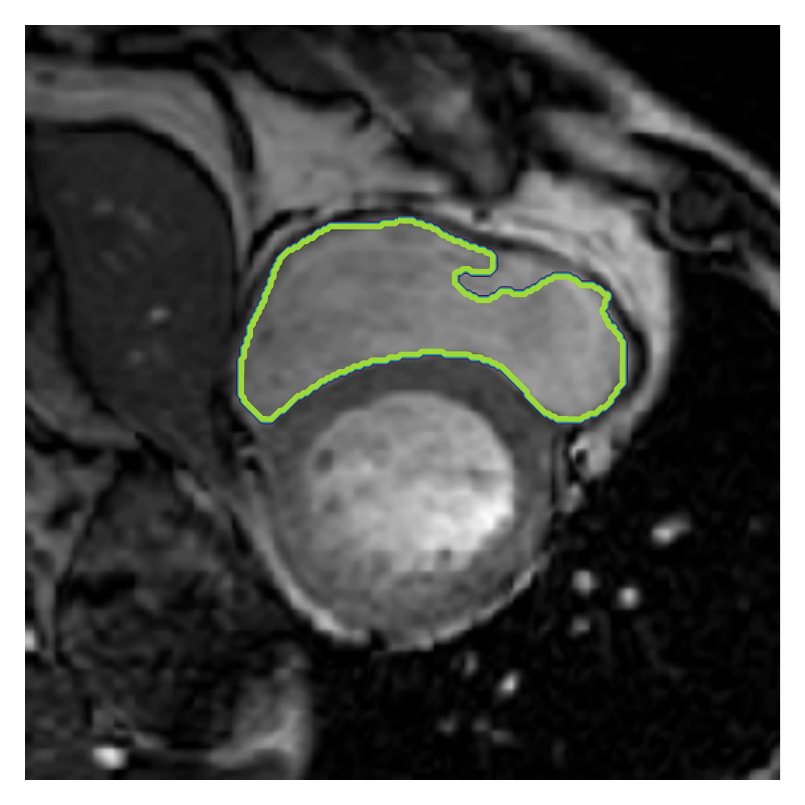} &
         \includegraphics[width=0.195\columnwidth, trim=1.7cm 1.3cm 1.3cm 1.7cm, clip]{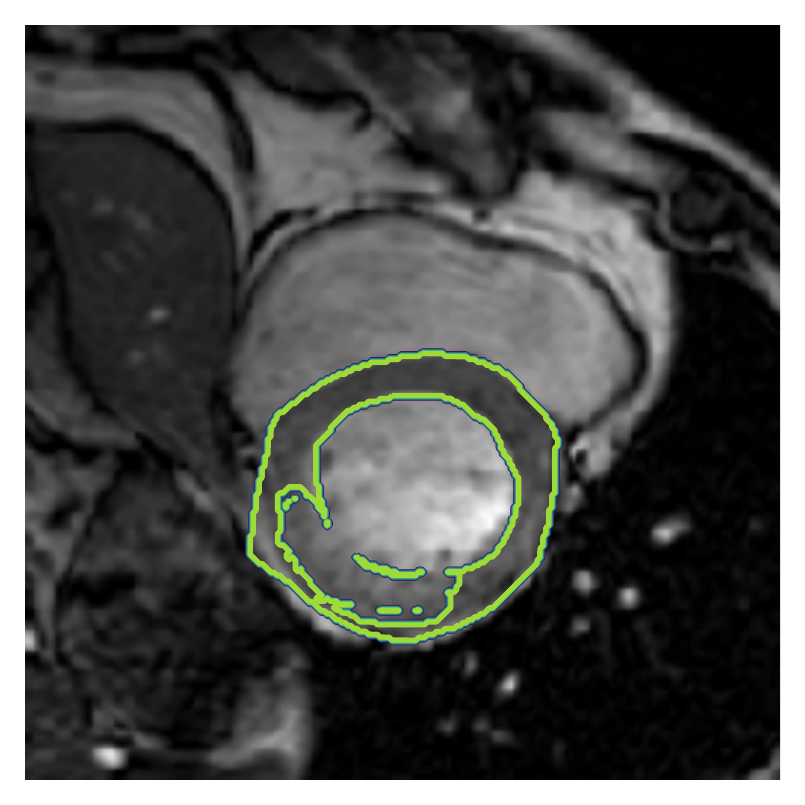} &
         \includegraphics[width=0.195\columnwidth, trim=1.7cm 1.3cm 1.3cm 1.7cm, clip]{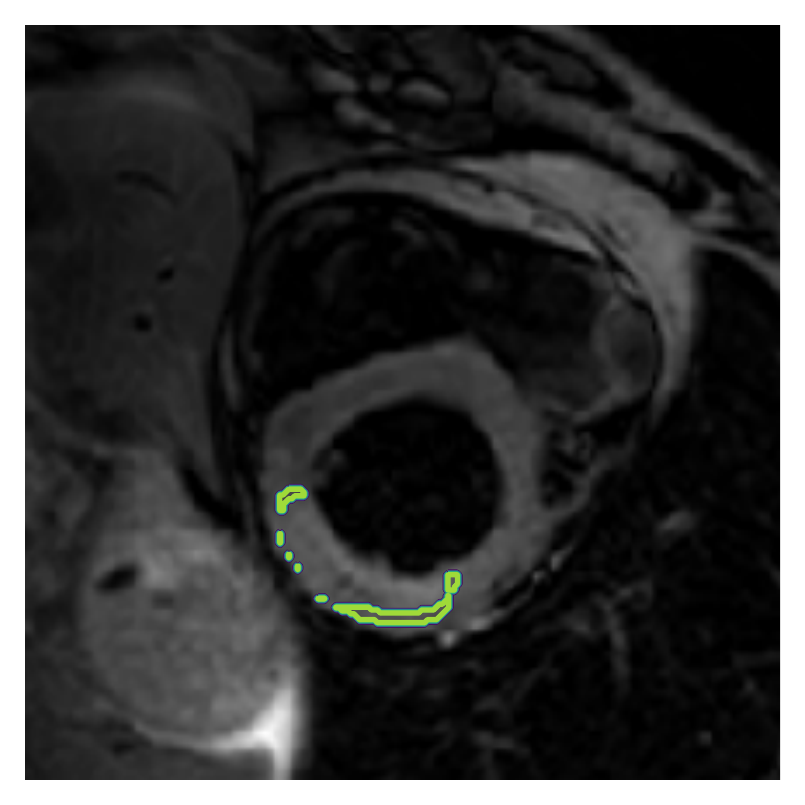} &
         \includegraphics[width=0.195\columnwidth, trim=1.7cm 1.3cm 1.3cm 1.7cm, clip]{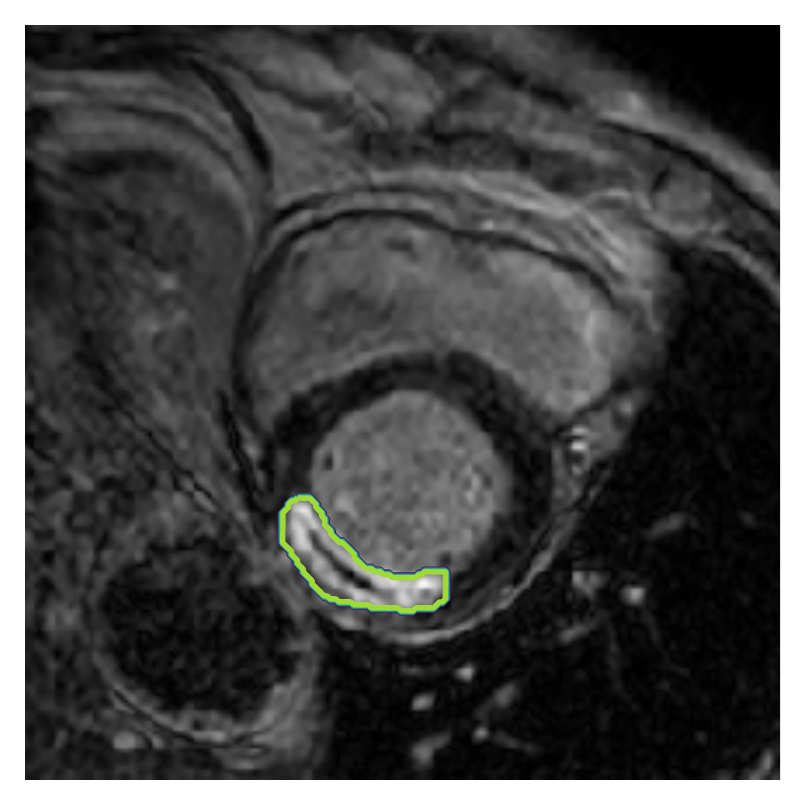} \\
         (a) LV BG & (b) RV BG & (c) LV NM & (d) LV ME & (e) LV MS\\
    \end{tabular}}
    \caption{Examples showing ground truth and predicted contours where the proposed method had achieved the highest dice score for LV ME+MS ($75.1\%$) and LV MS ($82.3\%$) with the manual delineations in the test set. The first three columns show the predicted contours against the bSSFP and the fourth and last columns show the predicted contours against T2 and LGE respectively. The rows correspond to different slices in the best case.}
    \label{fig:best}
\end{figure}

\begin{figure}[h!]
    \centering
    \setlength{\tabcolsep}{1pt}
    {\scriptsize
    \begin{tabular}{ccccc}
         \includegraphics[width=0.195\columnwidth, trim=2.3cm 0.5cm 2.5cm 4.0cm, clip]{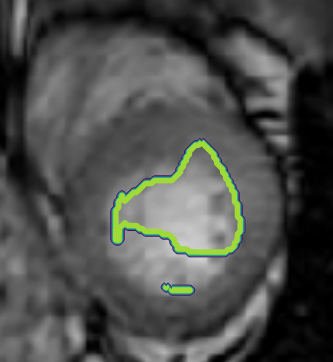} &
         \includegraphics[width=0.195\columnwidth, trim=2.3cm 0.5cm 2.5cm 4.0cm, clip]{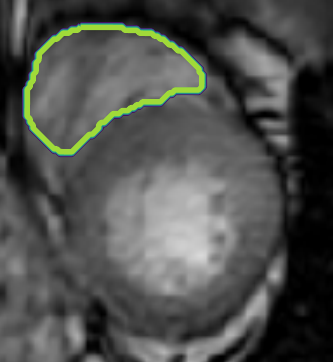} &
         \includegraphics[width=0.195\columnwidth, trim=2.3cm 0.5cm 2.5cm 4.0cm, clip]{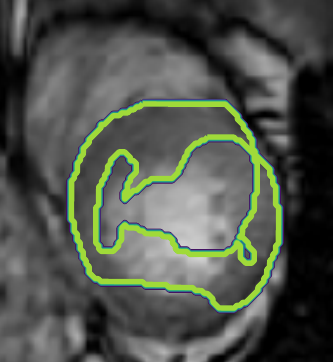} &
         \includegraphics[width=0.195\columnwidth, trim=2.3cm 0.5cm 2.5cm 4.0cm, clip]{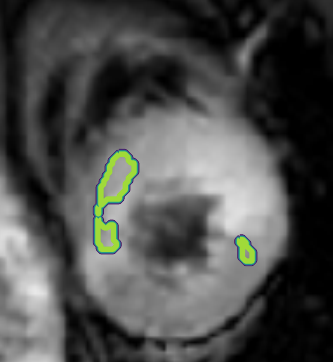} &
         \includegraphics[width=0.195\columnwidth, trim=2.3cm 0.5cm 2.5cm 4.0cm, clip]{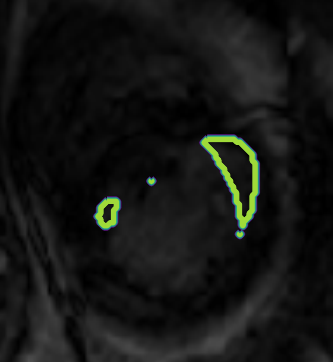} \\
         \includegraphics[width=0.195\columnwidth, trim=2cm 0.2cm 2cm 3.2cm, clip]{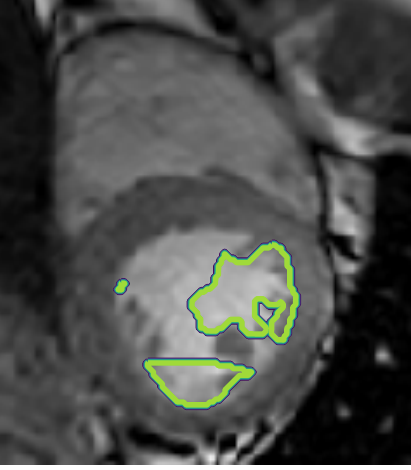} &
         \includegraphics[width=0.195\columnwidth, trim=2cm 0.2cm 2cm 3.2cm, clip]{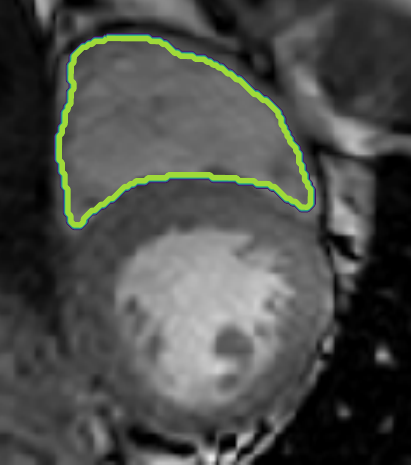} &
         \includegraphics[width=0.195\columnwidth, trim=2cm 0.2cm 2cm 3.2cm, clip]{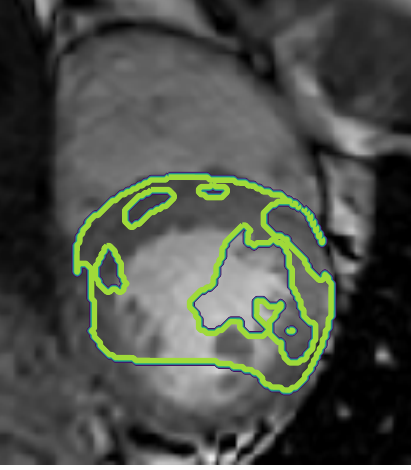} &
         \includegraphics[width=0.195\columnwidth, trim=2cm 0.2cm 2cm 3.2cm, clip]{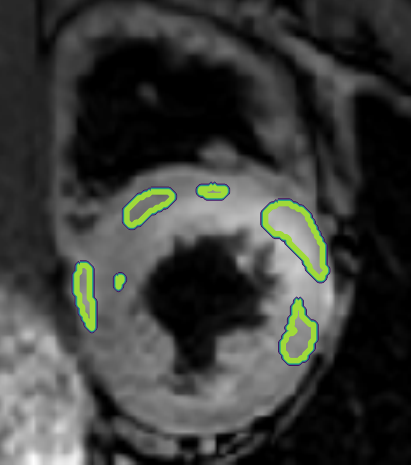} &
         \includegraphics[width=0.195\columnwidth, trim=2cm 0.2cm 2cm 3.2cm, clip]{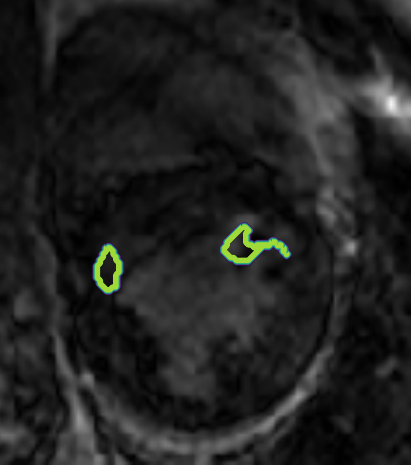} \\
         \includegraphics[width=0.195\columnwidth, trim=2cm 0.1cm 1.8cm 2.7cm, clip]{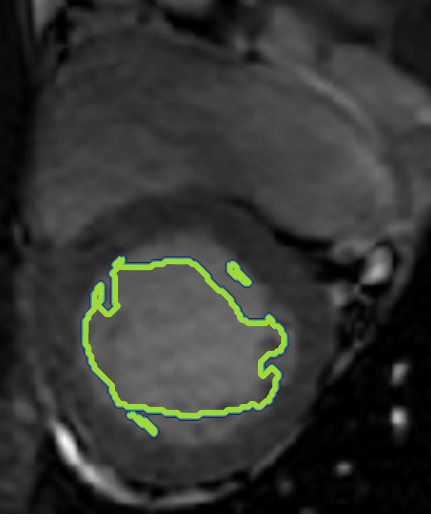} &
         \includegraphics[width=0.195\columnwidth, trim=2cm 0.1cm 1.8cm 2.7cm, clip]{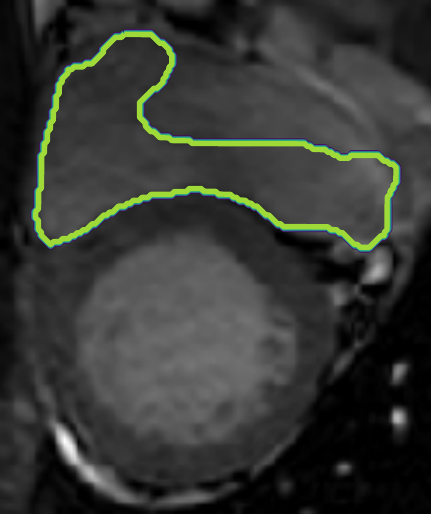} &
         \includegraphics[width=0.195\columnwidth, trim=2cm 0.1cm 1.8cm 2.7cm, clip]{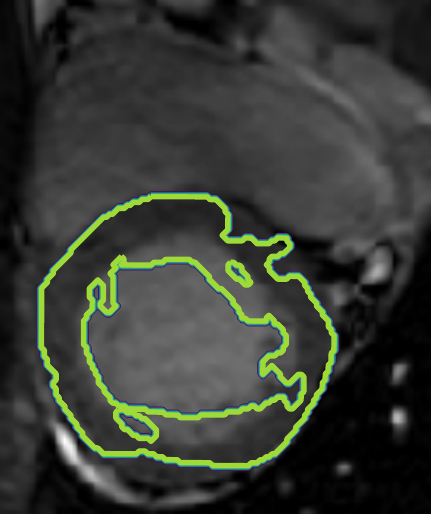} &
         \includegraphics[width=0.195\columnwidth, trim=2cm 0.1cm 1.8cm 2.7cm, clip]{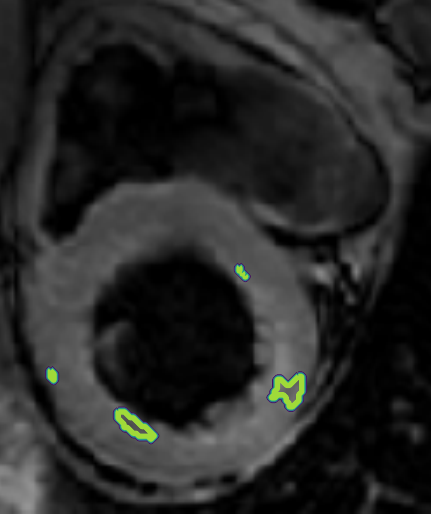} &
         \includegraphics[width=0.195\columnwidth, trim=2cm 0.1cm 1.8cm 2.7cm, clip]{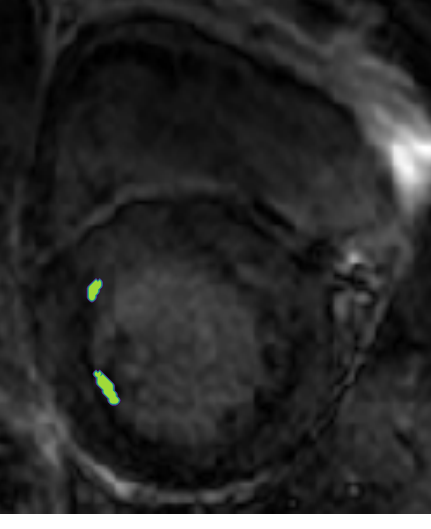} \\
         (a) LV BG & (b) RV BG & (c) LV NM & (d) LV ME & (e) LV MS\\
    \end{tabular}}
    \caption{Examples showing ground truth and predicted contours where the proposed method had achieved the lowest dice score for LV ME+MS ($0.1\%$) and LV MS ($18.1\%$) with the manual delineations in the test set. The first three columns show the predicted contours against the bSSFP and the fourth and last columns show the predicted contours against T2 and LGE respectively. The rows correspond to different slices in the worst case.}
    \label{fig:worst}
\end{figure}

We select the case that achieves the highest and lowest dice score for visual assessment. Fig.~\ref{fig:best} shows example segmentation results where the proposed method achieved the highest agreement with the ground truth delineations. Fig.~\ref{fig:worst} shows example segmentation results where the proposed method achieved the lowest agreement with the ground truth delineations.

\section{Conclusion}
We proposed a fully automated approach to segment the LV ME and LV MS from multi-sequence CMR data. We introduce an augmentation module to enhance the training set and a linear encoder and decoder along with a network module to improve the segmentation performance. The algorithm is trained using the 25 cases provided by the challenge and the evaluation is performed by the challenge organizers on another 20 cases which are not included in the training set. The proposed method yielded a overall mean dice metric of $46.8\%$, $55.7\%$ for LV ME and LV ME+MS delineations.

\section{Acknowledgment}
The authors wish to thank the challenge organizers for providing training and test datasets as well as performing the algorithm evaluation. The authors of this paper declare that the segmentation method they implemented for participation in the MyoPS 2020 challenge has not used additional MRI datasets other than those provided by the organizers. The authors would wish to acknowledge Compute Canada for providing the computation resource.

\bibliographystyle{splncs04}
\bibliography{ref}

\end{document}